\documentclass[12pt]{amsart}
\usepackage{graphicx,amsmath,amssymb}

\input epsf
\newcommand{\C}{\mathbb C}
\newcommand{\R}{\mathbb R}

\renewcommand{\d}{\prime} 
\newcommand{\dd}{{\prime \prime}}

\renewcommand{\Re}{{\rm Re}\,}
\renewcommand{\Im}{{\rm Im}\,}
\newtheorem{theorem}{Theorem}
\newtheorem{lemma}[theorem]{Lemma}
 
\newtheorem{corollary}[theorem]{Corollary}

\newtheorem{conjecture}{Conjecture}

\newtheorem{remark}{Remark}

\newtheorem{definition}{Definition}

\begin{document}
\title{On the Eigenproblems of $\mathcal {PT}$-Symmetric Oscillators} 

\author{K. C. Shin}
\address{Department of Mathematics, University of Illinois, Urbana, IL 61801}
\date{July 3, 2000}

\begin{abstract}
We consider the non-Hermitian Hamiltonian $H=-\frac{d^2}{d x^2}+P(x^2) -
(ix)^{2n+1}$ on the real line, where  $P(x)$ is a polynomial of
degree at most $n \geq 1$  with all nonnegative real coefficients (possibly
$P \equiv 0$).  
It is proved that the eigenvalues $\lambda$ must be in
the sector $|\arg \lambda| \leq \frac{\pi}{2n+3}$.  
 Also for the case $H=-\frac{d^2}{d x^2}-(ix)^3$ , we establish a zero-free
region of the eigenfunction $u$ and its derivative $u^{\d}$
and we find some other interesting properties of eigenfunctions.   
\end{abstract}

\maketitle

\begin{center}
{\it Preprint.}
\end{center}

\baselineskip = 18pt

\section{Introduction}
\label{introduction}
\vskip 12pt
We are considering the eigenproblem 
\begin{equation} \label{eigen}
- u^\dd(x) + [P(x^2) -(ix)^{2n+1}] u(x) = \lambda u(x) \qquad
\text{for $-\infty < x < \infty$} 
\end{equation} 
with $u(\pm \infty)=0$, where $P(x)$ is a polynomial of
degree at most $n \geq 1$ with all
nonnegative real coefficients (possibly $P \equiv 0$).

This is an  example of a class of problems, the so-called
$\mathcal {PT}$-symmetric non-Hermitian Hamiltonian problems, which
have arisen in recent years in a number of physical contexts \cite{HN1,HN2,NS}.
D. Bessis conjectured in 1995 that:
\begin{conjecture}
 Eigenvalues of  $H=-\frac{d^2}{d x^2}-(ix)^3$ are all 
real and positive.
\end{conjecture}
  Many numerical and asymptotic results 
\cite{BBM,BCMS,BD,BJK}  support this conjecture. 
And later for $n >1$ it was conjectured that the equation
(\ref{eigen}) also
has positive real eigenvalues, under different boundary conditions
\cite{BB}. However, there is no rigorous proof of this to date.

This paper is organized as follows: In Section \ref{sector}, we prove that
eigenvalues of the equation (\ref {eigen}) lie in the 
sector $|\arg \lambda| \leq \frac{\pi}{2n+3}$. This goes part way to
proving that the eigenvalues are real and positive. We generalize this
result to $H=-\frac{d^2}{d x^2}+ [P(x^2)+ixQ(x^2)]$ for  some real
polynomials $P$ and $Q$. In particular, for the potentials 
$-(ix)^3$ and $x^2+i g x^3$ with any real $g$, we have that $|\arg \lambda|\leq \frac{\pi}{5}$. Then next in Section 
\ref{zero-free}, for the case $H=-\frac{d^2}{d x^2}-(ix)^3$, we
fairly precisely locate the zeros of the  eigenfunctions and their first
derivatives in the complex plane. Conversely we find a large zero-free
region. In Section \ref{other}, still with $H=-\frac{d^2}{d
x^2}-(ix)^3$, we find a large  class of polynomials that are  
orthogonal to $|u|^2$ on each horizontal line. And finally in the last
section, we discuss related open problems.

For the rest of Introduction, we provide some more background
information on (\ref{eigen}). First, a $\mathcal {PT}$-symmetric
Hamiltonian is a Hamiltonian which is invariant under the product of
the parity operation $\mathcal P(: x \mapsto -x)$
and the time reversal operation $\mathcal T(: i \mapsto
-i)$. Certainly (\ref{eigen}) is $\mathcal {PT}$-symmetric while, for
example, $-\frac{d^2}{d x^2}+x-(ix)^3 $ is not $\mathcal
{PT}$-symmetric. If $H=-\frac{d^2}{d x^2}+V(x)$ is
$\mathcal{PT}$-symmetric, then $\overline {V(-x)}=V(x)$ and so $\Re V(x)$ is
an even function and $\Im V(x)$ is an odd function. Hence if $V(x)$ is
a polynomial, then $V(x) =P(x^2)+ixQ(x^2)$ for some real polynomials
$P$ and $Q$.  

Next by the work of Caliceti {\it et al.} \cite{C, CGM}, it is known that the 
$\mathcal {PT}$-symmetric
Hamiltonian $H=-\frac{d^2}{d x^2}+x^2-g(ix)^3$ has discrete spectrum, for $g$ real, and these eigenvalues are positive real if $g$ is small enough. 
However, there are some $\mathcal {PT}$-symmetric Hamiltonians that have
no eigenvalues \cite[\S1]{M}, or non-real eigenvalues \cite[footnote
on page 26]{DP}.   

Lastly, for any $\lambda \in \C$ there are two linearly independent
solutions of (\ref {eigen}), if the boundary conditions are not
imposed. In
generic cases, the solutions   blow up at both $+\infty$ and
$-\infty$, while in exceptional cases, the solutions decay to zero  
as $x$ approaches $+ \infty$ or $-\infty$. Only in very exceptional
cases (when $\lambda$ is an eigenvalue!) does one find a solution that
decays  to zero at both $+\infty$ and $-\infty$ (see Lemma \ref{asymp}
for details). 
\section{The eigenvalues lie in a sector}
\label{sector}
In this section, we prove that the eigenvalues $\lambda$ of
(\ref{eigen}) lie in the sector $|\arg \lambda| \leq \frac{\pi}{2n+3}$
and we extend this result for more general cases. To do this we will use
 results of Hille \cite[\S7.4]{H}.

For any $\lambda \in \C$ the equation (\ref {eigen}) without
the boundary conditions allows two linearly independent
solutions. If $u(x)$ solves the ODE (\ref{eigen}), then since
$P(z^2) -(iz)^{2n+1}$ is an entire function (analytic in the whole complex
plane), there 
exists an entire function $u(z)$  which agrees with $u(x)$ on the real
line and  satisfies $- u^\dd(z) + [P(z^2) -(iz)^{2n+1}] u(z) =
\lambda u(z) $. We begin by describing the asymptotic behavior of $u$ near infinity. Recall that $\deg P \leq n$.  
\begin{definition}\label{angle}
{\rm Let 
\begin{eqnarray}
\theta_j =  2\pi\frac{j}{2n+3}-\frac{\arg (i^{2n+1})}{2n+3}
         = \left\{
            \begin{array}{rl}
            \frac{2\pi j-\frac{\pi}{2}}{2n+3} & \quad \text{if $n$ is even,}\\
            \frac{2\pi j+\frac{\pi}{2}}{2n+3} & \quad \text{if $n$ is odd.}
            \end{array}
            \right. \nonumber
\end{eqnarray}

 We define {\it Stokes regions} 
$$S_j = \{ z \in \C:\theta_j < \arg z
< \theta_{j+1}\},$$ for $j= 0, 1, 2,
..., 2n+2$. And for notational convenience, we define $S_{j+2n+3} = S_j$
for all $j$.
 Also we  denote
$$S_{j,\epsilon} = \{ z \in \C:\theta_j +\epsilon < \arg z
<\theta_{j+1} -\epsilon\} ,$$ for $0< \epsilon < \frac{\pi}{2n+3}.$}
\end{definition}
Notice  $\theta_j$ is neither $0$ nor $\pi$. Thus the negative and the 
positive 
real axes lie within two of the Stokes regions (see Figure 1). We call these  
the {\it left-} and the {\it right-hand} Stokes regions, respectively. Also we call 
the rays $\{\arg z=\theta_j\}$ ``critical rays''.
\begin{figure}[t]
    \begin{center}
    \includegraphics[width=.2\textwidth]{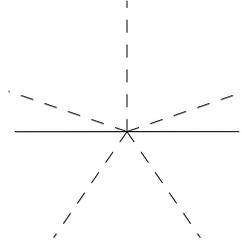}
    \end{center}
 \vspace{-.5cm}
\caption{For $n=1$; the solid line is the real axis and the dotted rays
are the critical rays, $\arg z=\theta_j= \frac{\pi}{10},
\frac{\pi}{2},\frac{9\pi}{10},  \frac{13\pi}{10}$ and $
\frac{17\pi}{10}$. } 
\end{figure}
\begin{lemma} \label{asymp}
Every solution of $- u^\dd(z) + [P(z^2)
-(iz)^{2n+1}] u(z) = \lambda u(z)$ is asymptotic to 
\begin{equation} \label{asymp-formula}
(const.)z^{-\frac{2n+1}{4}}\exp\left[{\pm \frac{2}{2n+3}
(iz)^{\frac{2n+3}{2}}(1+o(1))}\right] 
\end{equation}
as $z \rightarrow \infty$ in $S_{j,\epsilon}$, for each $0< \epsilon <
\frac{\pi}{2n+3}.$ The error $o(1)$ is uniform in $\arg z$ in the
sense that  $\lim_{r \rightarrow \infty} \sup \{|o(1)|: z \in S_{j,\epsilon},
 |z|=r\} =0$. 

Also $u$ has infinitely many zeros in $\C$ but only finitely many in 
$\cup_j S_{j,\epsilon},$ for each $0 < \epsilon < \frac{\pi}{2n+3}$.
\end{lemma}
The asymptotic expressions imply in particular that in each Stokes
region, $u(z)$ either decays to 0 or blows up, as $z$ approaches infinity in  
$S_{j,\epsilon}$. 
\begin{proof}
See Hille's book \cite[\S7.4]{H} for a proof of a more general result. An outline of the proof
is as follows: 
Hille first  transforms the equation into another complex $Z$-plane by using
the Liouville transform. And then he compares $u$ with the solutions of the
sine equation $w^\dd(Z) +w(Z)=0$ and finally transforms back to the
original complex $z$-plane. So the above asymptotic expressions are
the asymptotic  
expressions for solutions of the sine equation (in the $Z$-variable)
expressed in terms of the original $z$-variable. The Stokes regions are
determined by the Liouville transformation. 

Also we can deduce the last assertion of the theorem from \cite[\S7.4]{H}. 
This is proved in \cite[Theorem 5]{G} for more general equations. 
\end{proof}
\begin{remark}{\rm Under the Liouville transformation, a neighborhood
of infinity in each 
Stokes region in the complex 
$z$-plane maps to a neighborhood of infinity in either the upper or
lower half $Z$-plane. So if 
$u$ decays in a Stokes region $S_{j}$ for some $j$, then $u$ must
blow up in the Stokes regions $S_{j+1}$ and $S_{j-1}$. Otherwise,
there would be a solution of the sine equation in the $Z$-plane which
decays to zero in all directions. This is a contradiction. However, $u$ 
might blow up in many consecutive Stokes regions 
(even in all Stokes regions) (see \cite[\S7.4]{H}).}
\end{remark}
\begin{definition} \label{eigen-defn}
{\rm Let $\lambda \in \C$ and let $u(z) \not \equiv 0$ be an analytic
function on 
$\C$ that satisfies (\ref{eigen}). We say $u$ is an {\it eigenfunction} and
$\lambda$ is an {\it eigenvalue}, for (\ref{eigen}), 
if $u(z)$ decays to zero along rays to infinity in the left- and
right-hand Stokes regions (that is, if $u$ has decaying 
asymptotics in (\ref{asymp-formula}), in these two regions).} 
\end{definition}
\begin{remark}\label{simple}
{\rm Given a Stokes region $S_j$, there always exists a solution of 
$-u^\dd(z) + [P(z^2)-(iz)^{2n+1}] u(z) = \lambda u(z)$ that blows up in $S_j$
\cite[\S7.4]{H}. So if there were two linearly independent  eigenfunctions 
with the same eigenvalue, then all the solutions of $-u^\dd(z) +  [P(z^2)
-(iz)^{2n+1}] u(z) = \lambda u(z)$ would satisfy $u(\pm
\infty+0i)=0$ and there would be no solutions that blow up in the left- and right-hand Stokes regions.  
Thus there are no repeated eigenvalues, and all eigenvalues are simple.} 
\end{remark}

\begin{remark}\label{conjugate}
{\rm Note that if $u(z)$ is an eigenfunction with  eigenvalue
$\lambda$, then $\bar u(-\bar z)$ is an eigenfunction with 
eigenvalue $\bar \lambda$ (an upper bar denotes the complex
conjugate). So if  an eigenvalue is real then $u(z)=c\bar u(- \bar z)$
by Remark \ref{simple}, and clearly $|c|=1$. Writing $c=e^{-2i\phi}$ and replacing $u$ by $e^{i\phi}u$, we get that eigenfunctions with real eigenvalues are
symmetric with respect to the imaginary axis.} 
\end{remark}
The main result of this paper is:
\begin{theorem} \label{sector-theorem}
If $\lambda$ is an eigenvalue of (\ref{eigen}), then $\lambda \not =0$
and $|\arg \lambda|
\leq \frac{\pi}{2n+3}$. 
\end{theorem}
That the eigenvalues have positive real part was known already
\cite{K} (according to Mezincescu \cite{M}); our proof below includes a very simple argument for
this fact. In the proof and elsewhere, we will use the following:

Since $u(z)$ decays exponentially along rays to infinity in the
left- and right-hand Stokes regions, so does $u^\d$ by the Cauchy integral 
formula. Therefore $p(r)|u(re^{i\theta})|^2$ and 
$p(r)|u^\d(re^{i\theta})|^2$ are integrable 
along the line $r \mapsto re^{i\theta}$ in $\C$ for any polynomial
$p(r)$, provided $|\theta|< \frac{\pi}{2(2n+3)}$ (so that the ends of the line stay in the left- and right-hand Stokes regions).
\begin{proof}[Proof of Theorem ~\ref{sector-theorem}]
Let $u$ be an eigenfunction with eigenvalue $\lambda$, so that
$$
u^\dd(z)+[-P(z^2) +(iz)^{2n+1}]u(z)= -\lambda u(z), 
$$ 
where $P(z)= 
\Sigma_{k=0}^{n} a_k z^k$ for some $a_k \geq 0$, $k=0, 1, 2, ..., n$.

Write 
$$\lambda=\alpha + i\beta,\quad \alpha , \beta \in \R.$$
Fix $\theta$ with $|\theta|< \frac{\pi}{2(2n+3)}$. 
Let $v(r)=u(re^{i\theta})$. Then $v^\d(r)=u^\d(re^{i\theta})e^{i\theta}$
and $v^\dd(r)=u^\dd(re^{i\theta})e^{2i\theta}.$
Thus our ODE becomes
$$
v^\dd(r)+\left\{[\alpha+i\beta-P(r^2e^{2i\theta})]e^{2i\theta}+
i^{2n+1}r^{2n+1}e^{i(2n+3)\theta}\right\} v(r)=0. 
$$   
Then we multiply this by $e^{-i(2n+3)\theta}\bar v(r)$, integrate  and use 
integration by parts to get 
\begin{eqnarray}
& & e^{-i(2n+3)\theta}\int_{-\infty}^{\infty}|v^\d|^2 dr \label{integral1}\\
&=&(\alpha
+i\beta)e^{-i(2n+1)\theta}\int_{-\infty}^{\infty}|v|^2 dr-
\int_{-\infty}^{\infty} e^{-i(2n+1)\theta}P(r^2e^{2i\theta}) |v|^2 dr
+i^{2n+1}\int_{-\infty}^{\infty}r^{2n+1}|v|^2 dr,\nonumber 
\end{eqnarray}
for all $|\theta| < \frac{\pi}{2(2n+3)}$, where we note that the line 
$re^{i\theta}$ stays in the left- and right-hand Stokes regions where 
$u$ (and hence $u^\d$) decays exponentially to zero as $z$ approaches
infinity.   

Taking the real part of (\ref{integral1}) gives (since $|\theta|<
\frac{\pi}{2(2n+3)}$) 
\begin{eqnarray}
  0 &< &\cos(2n+3)\theta\int_{-\infty}^{\infty} |v^\d|^2 dr \label{integral2}\\
& = &\left\{\alpha\cos(2n+1)\theta 
+\beta\sin(2n+1)\theta\right\}\int_{-\infty}^{\infty} |v|^2 dr
\nonumber\\ 
& &-\int_{-\infty}^{\infty}\Re [e^{-i(2n+1)\theta}P(r^2e^{2i\theta})]
|v|^2 dr.\nonumber 
\end{eqnarray}
But $\Re [e^{-i(2n+1)\theta}P(r^2e^{2i\theta})] = \Sigma_{k=0}^{n} a_k r^{2k}
\cos(2n-2k+1)\theta \geq 0$ if $a_k \geq 0$ and
$|\theta|<\frac{\pi}{2(2n+1)}$ (certainly true if
$|\theta|<\frac{\pi}{2(2n+3)}$). 
So from (\ref{integral2}) we conclude that
\begin{equation} \nonumber
\alpha \cos(2n+1)\theta+\beta \sin(2n+1)\theta > 0,
\end{equation}
for all $|\theta| < \frac{\pi}{2(2n+3)}$. That is,
$$
\alpha > |\beta| \tan(2n+1)\theta,
$$
for all $0 \leq \theta < \frac{\pi}{2(2n+3)}.$

Taking $\theta=0$ gives $\alpha>0$, in particular $\lambda \not =0$ and 
 taking $\theta \rightarrow \frac{\pi}{2(2n+3)}$ gives 
$$\alpha \geq |\beta| \tan\frac{(2n+1)\pi}{2(2n+3)}.$$
Then finally using $\tan \phi = \cot (\frac{\pi}{2}-\phi)$, we have 
$$ \tan\frac{\pi}{2n+3} \geq \frac{|\beta|}{\alpha}.$$
That is, $|\arg \lambda| \leq \frac{\pi}{2n+3}.$
\end{proof}
\begin{remark}\label{general}
{\rm We can extend Theorem \ref{sector-theorem} by allowing $P$ to have some
negative coefficients  as long as $P$ satisfies  $\Re
[e^{-i(2n+1)\theta}P(r^2e^{2i\theta})] \geq 0$ 
for $|\theta| < \frac{\pi}{2(2n+3)}$. For example, with
$n=3$ and  $c \in \R$, let $P(z)=z^3+cz^2+z$; then $\Re
[e^{-7i\theta}P(r^2e^{2i\theta})]=r^2[r^4
\cos\theta+cr^2\cos(3\theta)+ \cos(5\theta)]$. So if
$c^2\cos^2(3\theta)-4\cos\theta\cos(5\theta) \leq 0$ for $|\theta| <
\frac{\pi}{18}$,  i.e. $|c|\leq
\sqrt{\frac{16}{3}\cos(\frac{\pi}{18})\cos(\frac{5\pi}{18})}$ $\approx
1.837$,  then $\Re [e^{-7i\theta}P(r^2e^{2i\theta})] \geq 0$. So the
theorem holds for this $P$ provided  $c \geq
-\sqrt{\frac{16}{3}\cos(\frac{\pi}{18})\cos(\frac{5\pi}{18})}$. 
  
Also by simple change of variables, we get the same result for $H =
-\frac{d^2}{d z^2}+[P(z^2)-g(iz)^{2n+1}]$ for any non-zero real $g$.  

Moreover, by translations in $\C$, we have the same result for $H =
-\frac{d^2}{d z^2}+P((z-\xi)^2)-gi^{2n+1}(z-\xi)^{2n+1}$ for any $\xi
\in \C$.   For example, if $u$ solves $u^\dd(z)-iz^3 u(z)=-\lambda
u(z)$,  then $v(z)=u(z+ai)$ solves $v^\dd(z)+
[(3az^2-a^3)-iz(z^2-3a^2)]v(z)=-\lambda v(z)$ for any real number
$a$. Observe $v$ still satisfies the boundary conditions: $v(\pm \infty+0i)
=u(\pm \infty+ai)=0$.} 
\end{remark}
\begin{remark}
{\rm The readers should notice that our boundary conditions  
are different, for $n \geq 2$, from those Bender and Boettcher \cite{BB} take.
 In \cite{BB}, the zero boundary conditions of the
problems $-u^\dd-(iz)^N u=\lambda u$ for $N \geq 4$ are taken not on Stokes 
regions containing the real axis but instead on Stokes regions  which are 
near the negative imaginary axis for large $N$.}
\end{remark}

The next theorem extends Theorem \ref{sector-theorem}.
\begin{theorem}\label{extended_1}
Let $\lambda \in \C$ and $n\geq 1$. Suppose that $u$ solves the ODE
\begin{equation}\label{eigen-1}
u^\dd-[P(z^2)+izQ(z^2)]u=-\lambda u, \quad u(\pm \infty+0i  )=0,
\end{equation}
for some real polynomials $P(z)=\Sigma _{k=0}^{n} a_k z^k$ 
 and  $Q(z)=\Sigma_{k=0}^n b_k z^k$ with  all nonnegative 
$a_k$ and with $b_n \in \R-\{ 0\}$.  
If for all $k < n$ the coefficients $a_k, b_k$ satisfy 
\begin{equation}\label{sufficient_1}
\frac{\sin^2(2n-2k)\theta}{\cos(2n-2k+1)\theta \cos(2n-2k-1)\theta} b_k^2 
\leq \left\{
           \begin{array}{rl}
            4 a_k a_{k+1} & \text{if } n=1 \text{ and } k=0\\
            2 a_k a_{k+1} & \text{if } n>1 \text{ and } k=0, n-1\\
            a_k a_{k+1}   & \text{if } n>1 \text{ and } 1 \leq k \leq n-2
             \end{array}
      \right.       
\end{equation}
at $\theta=\frac{\pi}{2(2n+3)}$,
then $|\arg \lambda | \leq \frac{\pi}{2n+3}$.
\end{theorem}
 For $n=3$, the coefficients of $b_k^2$ in (\ref{sufficient_1}) are
 approximately $3.41, 0.74$ and  $0.14$ for $k=0, 1, 2,$
 respectively. 

Theorem \ref{extended_1} contains Theorem \ref{sector-theorem}, just
by taking $b_k=0$ for $k=0, 1, ..., n-1$ (in which case
(\ref{sufficient_1})  is trivially satisfied). 
\begin{proof}
The main idea of the proof is the same as that of the proof of 
Theorem \ref{sector-theorem}. Even if the equation (\ref{eigen-1}) is 
little different from the equation for (\ref{asymp-formula}), 
Stokes regions for (\ref{eigen-1}) 
are the same as for (\ref{asymp-formula}) (if $b_n$ has the same sign as 
$(-1)^{n+1}$)  or else are rotated by $180^\circ$ (if $b_n$ has the 
opposite sign). See Section $7.4$ in \cite{H} for details. But in either 
case, the lines $r \mapsto re^{i\theta}$ with $|\theta|< 
\frac{\pi}{2(2n+3)}$ lie within the left- and right-hand Stokes regions, 
where we impose the zero boundary conditions. And this gives the 
integrabilities in the proof.

Let $v(r)=u(re^{i\theta})$. Then like we derived (\ref{integral2}) 
in the proof of Theorem \ref{sector-theorem} we have
\begin{eqnarray}
 & &\left\{\alpha \cos(2n+1)\theta+\beta \sin(2n+1)\theta\right\}
\int_{-\infty}^{\infty} |v|^2 dr\nonumber\\ 
&= &\int_{-\infty}^{\infty} \left\{\cos[(2n+3)\theta]
|v^\d|^2+\Sigma_{k=0}^{n} a_k 
\cos[(2n-2k+1)\theta] r^{2k}|v|^2 \right.\label{sine_1}\\
& &\qquad + \left.\Sigma_{k=0}^{n-1} b_k \sin[(2n-2k)\theta]
r^{2k+1}|v|^2 \right\}dr, \quad \text{for  } |\theta|<
\frac{\pi}{2(2n+3)}.\nonumber 
\end{eqnarray}
Since $a_k\geq 0$ for all $k$,  we  get $\alpha >0$ by letting
$\theta=0$ in (\ref{sine_1}). (This is true for any $Q$ with real
coefficients and $\deg Q=n$.) 

If we find conditions on $a_k$ and $b_k$ such that
\begin{equation}\label{sine_2}
 \Sigma_{k=0}^{n} a_k\cos[(2n-2k+1)\theta] r^{2k}|v|^2+
\Sigma_{k=0}^{n-1} b_k \sin[(2n-2k)\theta] 
r^{2k+1}|v|^2  \geq 0,
\end{equation}
for every $r \in \R$ and $|\theta|<\frac{\pi}{2(2n+3)}$, then  we have from 
(\ref{sine_1}) with 
$\theta \rightarrow \pm \frac{\pi}{2(2n+3)}$, that 
$$\alpha \cos\frac{(2n+1)\pi}{2(2n+3)} \pm \beta
\sin\frac{(2n+1)\pi}{2(2n+3)} \geq 0.$$
 So then $|\arg \lambda| \leq \frac{\pi}{2n+3}$ as desired, like in
the proof of Theorem \ref{sector-theorem}.  

When $n \geq 3$, we can rewrite the expression in (\ref{sine_2}) as $|v|^2$ times 
\begin{eqnarray}
&&\qquad \qquad \bigl\{a_0 \cos(2n+1)\theta+r b_0 \sin2n\theta
+r^2 \frac{a_{1} }{2} \cos(2n-1)\theta \bigr\} 
\nonumber\\
&&+\Sigma_{k=1}^{n-2}\bigl\{\frac{a_k }{2} \cos(2n-2k+1)\theta+rb_k
\sin(2n-2k)\theta 
+r^2\frac{a_{k+1} }{2} \cos(2n-2k-1)\theta\bigr\}r^{2k} 
\label{sine_3}\\
&&\qquad \qquad \quad +\bigl\{\frac{a_{n-1}}{2} \cos3\theta+r b_{n-1}
\sin2\theta 
+r^2a_{n} \cos\theta\bigr\}r^{2n-2}.\nonumber
\end{eqnarray}

Now (\ref{sine_3}) is nonnegative if each quadratic in $r$ has
non-positive discriminant:  
\begin{eqnarray}
 b_0^2 \sin^2 2n\theta-2a_0a_1\cos(2n+1)\theta\cos(2n-1)\theta & \leq
& 0,\nonumber\\ 
b_k^2 \sin^2(2n-2k)\theta-a_k a_{k+1}
\cos(2n-2k+1)\theta\cos(2n-2k-1)\theta &\leq& 0 \quad \text{for} \quad
1 \leq k\leq n-2\nonumber,\\ 
 b_{n-1}^2 \sin^2 2\theta-2a_{n-1}a_n\cos3\theta\cos\theta & \leq & 0,\nonumber
\end{eqnarray} 
 which is (\ref{sufficient_1}). 
The coefficients of $b_k^2$ in (\ref{sufficient_1}) are all
increasing functions of $0\leq \theta<\frac{\pi}{2(2n+3)}$, and so it suffices that (\ref{sufficient_1}) hold at $\theta= \frac{\pi}{2(2n+3)}$.
 
Now when $n=1, 2$, it is easy to see similarly that the theorem holds. This completes the proof.
\end{proof}
\begin{remark}
{\rm In (\ref{sine_2}), the sign of $\int_{-\infty}^{\infty}
r^{2k+1}|v|^2 dr$ is difficult to determine because $r$ can be
negative as well as positive. 

The conditions in Theorem \ref{extended_1} are sufficient but not
necessary, as is clear from the proof.  

We used $a_k= \frac{a_k}{2} +  \frac{a_k}{2}$ to get (\ref{sine_3})
from (\ref{sine_1}). If we use $a_k=\delta_k a_k+(1-\delta_k)a_k$  for
some $0<\delta_k<1$, we will get new sufficient conditions for the
theorem.} 
\end{remark}
\section{The zero--free region for $u$ and $u^\d$}
\label{zero-free}
The results in
the previous section are based  on the eigenfunction $u$
decaying to zero as $z$ approaches infinity in the left- and
 the right-hand Stokes regions.   
So consideration of the {\it finite} zeros of $u$ may be useful
for further results on our eigenproblem. 

\begin{figure}[t]
    \begin{center}
    \includegraphics[width=.4\textwidth]{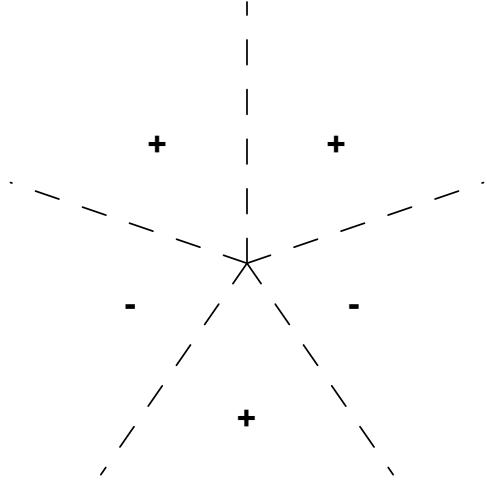}
    \end{center}
 \vspace{-.5cm}
\caption{In this Figure, rays are  $\arg z= \frac{\pi}{10},
\frac{\pi}{2},\frac{9\pi}{10}, \frac{13\pi}{10}$ and $
\frac{17\pi}{10}$. A ``$+$''  indicates that the eigenfunction is blowing
up while a ``$-$'' indicates that the eigenfunction is decaying to zero as
$z$  approaches infinity. }\label{f:shin2}
\end{figure}

For the next two sections, we will suppose $H=-\frac{d^2}{d z^2} -(iz)^3$.
See Figure \ref{f:shin2} for the asymptotic behavior of the eigenfunction $u$.
 
In this section, we provide a zero-free region for the eigenfunction
$u$ of 
\begin{equation}\label{eigen2}
u^\dd-iz^3u=-\lambda u \qquad \text{with $u(\pm \infty +0i)=0$}
\end{equation} 
and for its derivative $u^\d$. And we give some answers on how zeros of the
eigenfunction should be arranged in $\C$. 

It is obvious that $u$ and
$u^\d$ do not share a common zero. Otherwise, by (\ref{eigen2}), all
the derivatives of $u$ and $u$ itself would vanish at the zero, and so
$u  \equiv 0$. 

The following lemma is needed for our argument. Recall
$\lambda=\alpha+i\beta$.
\begin{lemma} \label{curve-lemma}
Let $z: [c, d]\rightarrow \C$ be a smooth curve with $z^\d(t) \not =0$ 
for $t \in [c, d]$. If $u$ solves (\ref{eigen2}), 
then writing $z(t)=x(t)+iy(t)$,  
\begin{eqnarray}
&   &\left.\Re(u^\d \bar u)
\right|_{z(c)}^{z(d)}  \label{real-curve} \\
&=& \int_c^d x^\d |u_x(z(t))|^2 dt+\int_c^d 
\left[x^\d \Re(iz^3(t)-\lambda)-y^\d \Im (iz^3(t)-\lambda) \right]|u(z(t))|^2
dt, \nonumber
\end{eqnarray}
and
\begin{eqnarray}
& &\left.\Im(u^\d \bar u )\right|_{z(c)}^{z(d)} \label{im-curve}\\
&=& -\int_c^d y^\d |u_x(z(t))|^2 dt +\int_c^d 
\left[y^\d \Re(iz^3(t)-\lambda)+x^\d \Im(iz^3(t)-\lambda )\right]|u(z(t))|^2
dt. \nonumber
\end{eqnarray}
\end{lemma}

Hille calls this lemma the Green's transform \cite[\S 11.3]{H}, and he uses it
 to get information on zero-free regions of solutions of linear second
 order equations (mainly with coefficient functions that are real on the real
 line). 
\begin{proof}
Let $f(t)=u(z(t))$ for $t \in [c,d]$. Then
$f^\d(t)=z^\d(t)u^\d (z(t))$ and 
$$
(\frac{f^\d(t)}{z^\d(t)})^\d=z^\d(t)u^\dd(z(t))=z^\d(t)[iz^3(t)-\lambda]f(t).
$$ 
 Hence by integration by parts,
$$
\left. \left(\frac{f^\d(t)}{z^\d(t)}\right) \bar f(t)\right|_c^d=\int_c^d
\frac{|f^\d|^2}{z^\d} dt+
\int_c^d z^\d[iz^3 -\lambda]|f|^2 dt.
$$
Now by the formula 
$f^\d(t)=z^\d(t)u^\d (z(t))$ and splitting real and imaginary
parts of the above, we get the lemma.
\end{proof}

Now we examine the consequences of the lemma. First,
 if  $\Re (u^\d \bar u)$ were not one-to-one on the
imaginary axis, that would imply that the eigenvalue would be real by
 (\ref{real-curve}) with $z(t)=it$.  
\begin{remark}\label{one-one}
{\rm Second, another immediate consequence of  Lemma \ref{curve-lemma} is 
that on any vertical line segments on which $\Im (iz^3-\lambda)$ doesn't 
change its sign, 
$\Re (u_x \bar u)$ as a function of $y$ is one-to-one. On horizontal 
line segments on which $\Im (iz^3-\lambda) $ doesn't change its sign, $\Im
(u_x \bar u)$ as a function of $x$ is one-to-one (Mezincescu
\cite[\S3]{M} observed this last fact on the real axis, where
$y\equiv 0$). These observations are special cases of Hille's Theorem 11.3.3
in \cite{H}.} 
\end{remark}
\begin{figure}[t]
    \begin{center}
    \includegraphics[width=.4\textwidth]{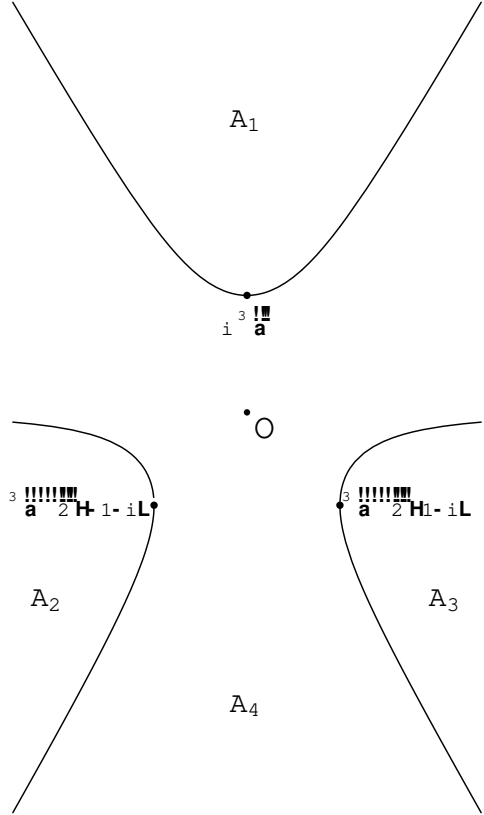}
    \end{center}
 \vspace{-.5cm}
\caption{ Here $\Re (iz^3-\lambda)$ is negative in $A_4$ since $\alpha>0$,
 while it is positive 
in $A_1, A_2$ and $A_3$.}\label{f:shin3} 
\end{figure}
\begin{figure}[t]
    \begin{center}
    \includegraphics[width=.4\textwidth]{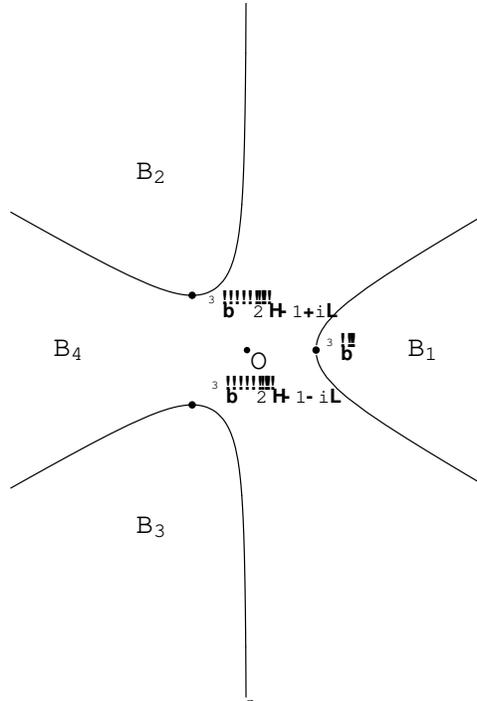}
    \end{center}
 \vspace{-.5cm}
\caption{ The level curves $\Im (iz^3-\lambda)=0$
with $\beta>0$ fixed. Here $\Im (iz^3-\lambda) $ is negative in $B_4$ while 
it is positive in $B_1, B_2$ and $B_3$.}\label{f:shin4} 
\end{figure}

\begin{figure}[t]
    \begin{center}
    \includegraphics[width=.4\textwidth]{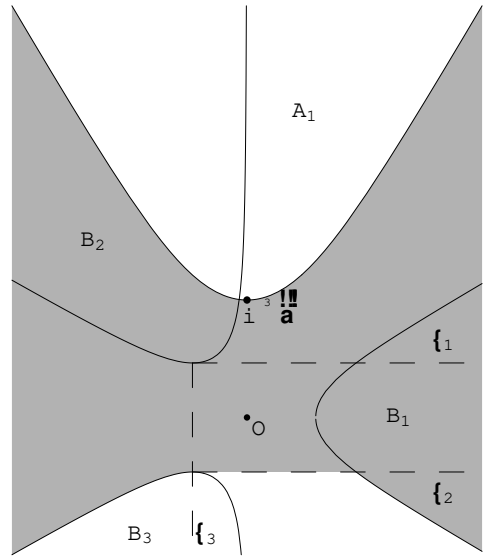}
    \end{center}
 \vspace{-.5cm}
\caption{$\Im (u^\d \bar u) < 0$ in the shaded area. Here $\ell_1: \Im z=
\sqrt[3]{\beta/2},$  $\ell_2: \Im z= -\sqrt[3]{\beta/2}$ and $\ell_3: \Re z=
-\sqrt[3]{\beta/2}$. }\label{f:shin5} 
\end{figure}

Third, let us define open regions $A_j$ and $B_j$, $j= 1, 2, 3, 4 $ as
in Figure \ref{f:shin3} and \ref{f:shin4}. The following two theorems
provide a large 
zero-free  region for an
eigenfunction $u$ of (\ref{eigen2}) and its derivative $u^\d$,
assuming $\lambda$ is non-real. Perhaps these theorems might help show that
$\lambda$ must actually be real. The underlying ideas of the proofs
are taken from Hille's book \cite[\S 11.3]{H}.
\begin{theorem} \label{im-part}
If $\beta:= \Im \lambda >0$ then $\Im (u^\d \bar u) < 0$  on 
$$ B_1 \cup \{z \in B_4: \Re z \leq
-\sqrt[3]{\beta/2}\} \cup \{z \not \in A_1: \Im z \geq
-\sqrt[3]{\beta/2}\}, \quad \text{see Figure \ref{f:shin5}.}$$
\end{theorem}
Mezincescu \cite[\S3]{M} has previously observed that the
eigenfunction has no zeros on the real axis, which obviously lies in the 
shaded region of Figure \ref{f:shin5}. Moreover, we see that all the zeros 
of $u$ and $ u^\d$ in $\Im z \geq 0$ must be in $A_1$.

Note that the lowest point in the closure $cl(B_2)$ of 
$B_2$ is $-\sqrt[3]{\beta/2}+ i\sqrt[3]{\beta/2}$.
\begin{proof}[Proof of Theorem ~\ref{im-part}]
For any $y \in \R$, by (\ref{im-curve}) with $z(t)=t+iy$ 
and by $u(\pm \infty+iy)= 0 =u^\d(\pm \infty+iy)$, we have that
$$
 \Im [u^\d(x+iy) \bar u(x+iy)]=\int_{-\infty}^x \Im
(iz(t)^3-\lambda)|u|^2 dt,
$$ 
and this is negative for $x+iy \in B_4$ with $|y|\leq
\sqrt[3]{\beta/2}$, because then $z(t)= t+iy \in B_4$ for all $ -\infty < t
< x$ and so $ \Im (iz(t)^3-\lambda)  <0$.

This argument also shows that $\Im (u^\d \bar u) < 0$ in $\{z
\in B_4: \Re z \leq  -\sqrt[3]{\beta/2}\}$; see Figure
\ref{f:shin5}. 

Similarly in $B_1$, for all $y$ we have  that
$ \Im (u^\d\bar u) =-\int_x^{\infty} \Im (iz(t)^3-\lambda) |u(t+iy)|^2
dt < 0$. 

For $ z \not \in A_1 \text{ with } \Im z \geq 
\sqrt[3]{\beta/2}$ (so that $z \in A_4$), we use (\ref {im-curve}) 
along vertical line
segments starting from points on the line $\Im z= \sqrt[3]{\beta/2}$
to conclude $\Im (u^\d\bar u) < 0$ in this region.  
\end{proof}

\begin{figure}
\noindent\begin{minipage}[c]{0.45\linewidth}
\centering
$$\epsfxsize = 2in \epsffile {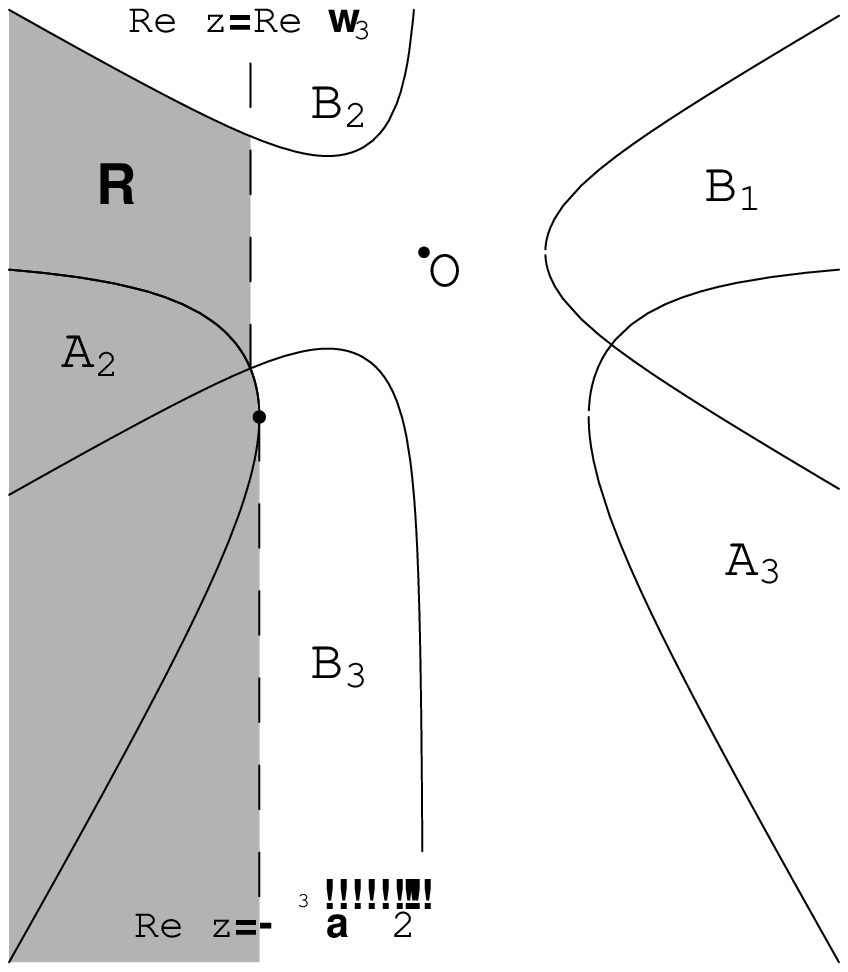}$$
\caption{$\Re (u^\d \bar u) > 0$ in the shaded area.} \label{f:shin6} 
\end{minipage}%
\noindent\begin{minipage}[c]{0.45\linewidth}
\centering
$$\epsfxsize = 2in \epsffile {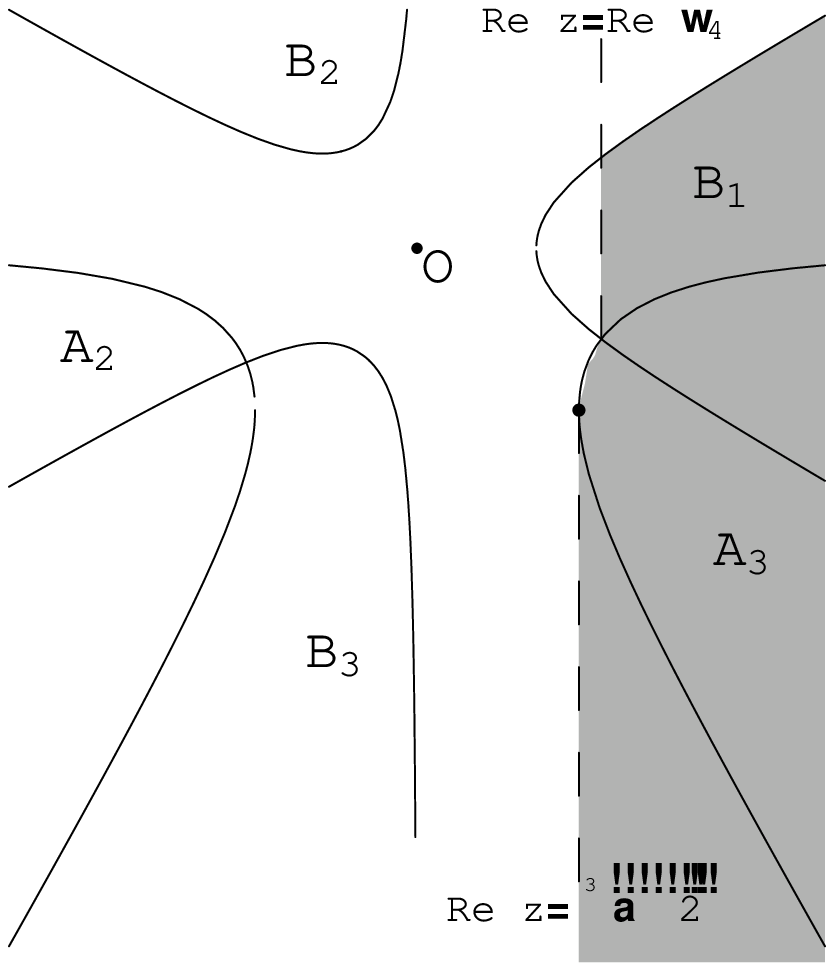}$$
\caption{$\Re (u^\d \bar u) < 0$ in the shaded area.} \label{f:shin7}
\end{minipage}
\end{figure}

Note that $ \Im (u^\d \bar u) = -\frac{1}{2}\frac{\partial}{\partial
y}|u(x+iy)|^2$. So in 
the region in Theorem \ref{im-part}, $|u(x+iy)|$ is an increasing function 
of $y.$  
\begin{theorem} \label{real-part}
Assume $\beta > 0$. Then
\item{(i)} 
$\Re (u^\d \bar u) > 0$ 
on the union of the regions $A_2$, the region below $A_2$ and the
region $\mathcal R \subset B_4$ between $A_2$ and $B_2$ with the real part less than or equal
to that of the zero $\omega_3$ of $iz^3-\lambda$ in the third quadrant. See
Figure \ref{f:shin6}.  
\item{(ii)} $\Re (u^\d \bar u) < 0$ on the union of the regions $A_3$,
the region below $ A_3$, and the region in $B_1$ with the real part
greater than or equal to that of the zero $\omega_4$ of $iz^3-\lambda $ in the
fourth  quadrant. See Figure \ref{f:shin7}. 
\end{theorem}
Obviously $iz^3-\lambda $ has  three zeros. When $\beta > 0$, one 
of the zeros is in the second quadrant,  one $\omega_3$ in the third
and one $\omega_4$ in the fourth quadrant. Certainly these are the three 
points at which the boundaries of the $A_i$ and $B_i$ intersect.

Theorem \ref{sector-theorem} with $n=1$ shows that 
$\frac{|\beta|}{\alpha} \leq \tan \frac{\pi}{5} \approx 0.73.$
It is easy to see that the rightmost point of $cl(A_2)$ is 
$\sqrt[3]{\frac{\alpha}{2}}(-1-i)$, at which 
$\Im (iz^3-\lambda)=x^3-3xy^2-\beta >0$ by $0<\beta <\alpha$. Thus the 
rightmost point of 
$cl(A_2)$ lies inside $B_3$ as shown in Figure \ref{f:shin6}. 
Similarly, the leftmost point of $cl(A_3)$ is  
$\sqrt[3]{\frac{\alpha}{2}}(1-i)$, at which 
$\Im (iz^3-\lambda)=x^3-3xy^2-\beta <0$. Thus  
the leftmost point of 
$cl(A_3)$ lies outside $B_1$ as shown in Figure \ref{f:shin7}.
\begin{proof}[Proof of Theorem ~\ref{real-part}]
 In the  regions $A_2$ and $A_3$, we use (\ref{real-curve}) with
horizontal lines to infinity to get the statements in parts $(i)$ and $(ii)$
of this theorem.   In the region $\mathcal R$ between $A_2$ and $B_2$ with  the
real part less than or equal to that of the zero of $iz^3-\lambda $ in
the third quadrant (see Figure \ref{f:shin6}),  we use (\ref {real-curve})
with vertical lines $z(t)=x+it$ to show that $\Re (u^\d\bar u) > 0$. 
That is, we use $\left. \Re (u^\d\bar
u)\right|_{x+ic}^{x+id}=-\int_c^d \Im (iz(t)^3-\lambda )|u(x+it)|^2 dt$. 
If $x+id \in \mathcal R$, we can find $x+ic \in cl(A_2 \cap B_3^{c})$,
 so that $\Re [u^\d(x+ic) \bar u(x+ic)] > 0$, and $-\Im (iz(t)^3-\lambda )>0$. 
Hence, the above integral is an increasing function of $d$. Hence we have 
the desired result in this region $\mathcal R$.

The region below $A_2$ is contained in $B_3$ since  the rightmost point 
of $cl(A_2)$ lies in $B_3$ (see Figure \ref{f:shin6}). So a similar argument  
shows that $\Re (u^\d\bar u) > 0$ in the region below $A_2$. 
Also, the region below $A_3$ is contained in $B_4$ (see Figure \ref{f:shin7})  
and so  modified arguments show that the other statements of this
theorem in part $(ii)$ are true. 
\end{proof}
\begin{corollary}
When $\Im \lambda=\beta >0$,
the zero-free region of $u$ and $u^\d$ contains the union of the three shaded regions in Figure \ref{f:shin5}, \ref{f:shin6} and \ref{f:shin7}. 
\end{corollary}

Note that in case $\Im \lambda=\beta < 0$ we can get similar theorems
corresponding to the above two, since $\bar u(-\bar z)$ is an
eigenfunction with eigenvalue $\bar \lambda$. The regions involved
are simply the reflections  of the above with respect to the imaginary
axis. 

\begin{figure}[t]
    \begin{center}
    \includegraphics[width=.4\textwidth]{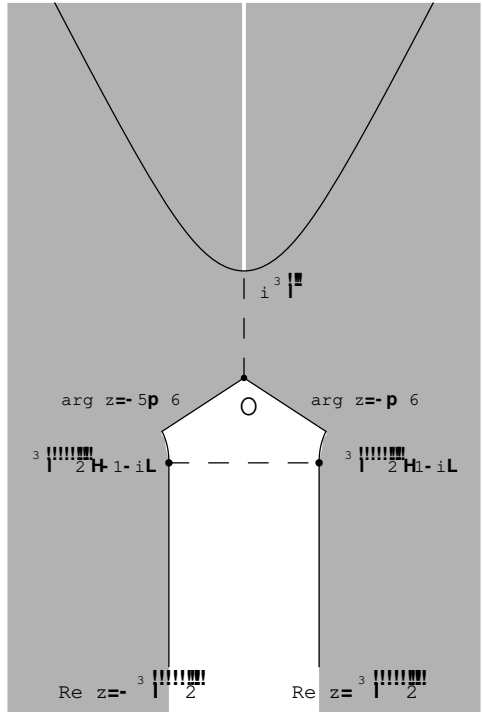}
    \end{center}
 \vspace{-.5cm}
\caption{When $\lambda$ is real, the shaded area is the zero-free region of $u$ and $u^\d$.}\label{f:shin11}
\end{figure}

In case $\beta=0$, so that $\lambda$ is real and $\lambda=\alpha>0$, the regions $B_1, B_2, B_3$ degenerate to the sectors $\{-\frac{\pi}{6} < \arg z <\frac{\pi}{6}\}, \{\frac{\pi}{2} < \arg z <\frac{5\pi}{6}\}, \{-\frac{5\pi}{6} < \arg z <-\frac{\pi}{2}\}$ respectively, and we get the following theorem on zero-free regions.
\begin{theorem}\label{realeigen}
Suppose $\lambda$ is real. Then $\Im (u^\d \bar u)<0$ on  $\{-\frac{\pi}{6} \leq \arg z \leq \frac{7\pi}{6}\} - A_1$ (which is a degenerate case of Figure \ref{f:shin5}), while $\Re (u^\d \bar u)$ behaves as in Figure \ref{f:shin6} and \ref{f:shin7} with $B_1, B_2, B_3$ being sectors as above.

Also  $\Re (u^\d \bar u)<0$ in $cl(A_1) \cap \{\Re z < 0\}$ and 
 $\Re (u^\d \bar u)>0$ in $cl(A_1) \cap \{\Re z >0\}$.
\end{theorem}
\begin{corollary}
When $\lambda$ is real, the zero-free region of $u$ and $u^\d$ contains the union of all regions in Theorem \ref{realeigen}; see Figure \ref{f:shin11}. That is, $u$ and $u^\d$ can only have zeros in $\{iy:y > \sqrt[3]{\lambda}\} $
$$
 \cup \left\{z \in A_4:-\frac{5\pi}{6}< \arg z <-\frac{\pi}{6}, \Im z > -\sqrt[3]{\lambda /2}\right\} \cup \left\{z:|\Re z|<\sqrt[3]{\lambda /2}, \Im z \leq  -\sqrt[3]{\lambda /2} \right\}.
$$
\end{corollary}
\begin{remark}
{\rm Bender {\it et al.} \cite{BBS} find numerically that $u$ has some zeros along an ``arch'' within the unshaded region in Figure \ref{f:shin11}, when $\lambda$ is real.}
\end{remark}

In proving Theorem \ref{realeigen}, we will use the following lemma.
\begin{lemma}\label{mono_sign2}
Suppose $\zeta \in A_1$,  
$\Re (u^\d \bar u)=0$ at $\zeta$. Then 
 $\Re (u^\d \bar u)<0$ at all $\zeta-t \in A_1, t>0$ and $\Re
(u^\d \bar u)>0$ at all $\zeta+t \in A_1, t>0$.
\end{lemma}
Note that there is no restriction on the sign of $\Im \lambda$, in this lemma.
\begin{proof}
 On horizontal line segments $z(t)=t+iy$ in $A_1$,
(\ref{real-curve}) becomes  
$$
\left.\Re(u^\d \bar u)\right|_{c+iy}^{d+iy}
= \int_c^d |u_x(z(t))|^2 dt+\int_c^d \Re(iz^3(t)-\lambda)|u(z(t))|^2 dt.
$$
Since $\Re(iz^3(t)-\lambda)>0 $ in $A_1$, $ \Re(u^\d \bar u)$ is a strictly 
increasing function of $x$ on each horizontal line segment in $A_1$. 
\end{proof}
\begin{proof}[Proof of Theorem ~\ref{realeigen}]
The proofs of Theorems \ref{im-part} and \ref{real-part} give everything except the last statement of the theorem. For that, recall we can take $u(z)=\bar u(-\bar z)$ by Remark \ref{conjugate}; this implies $\Re (u^\d \bar u)$ is an odd function with respect to reflection in the imaginary axis, so $\Re (u^\d \bar u)=0$ on the whole imaginary axis. Now we use Lemma \ref{mono_sign2} to complete the proof. 
\end{proof}

 By the last statement of Lemma \ref{asymp}, in the sector 
$S_{-1,\frac{\pi}{20}}$ that contains the negative imaginary axis, the 
eigenfunction $u$ has only finitely many 
zeros. Now with the zero-free region in the Theorems \ref{real-part} 
and \ref{realeigen}, we see that $u$ has only finitely many zeros 
in $\Im z <0$. 
Since $u$ has infinitely many zeros, $u$ must have infinitely many zeros in 
$\Im z \geq 0$. When $\beta >0$ (hence when $\beta<0$ as well), by 
Theorem \ref{im-part}, $u$ must have infinitely many zeros in $A_1$. Also when $\beta=0$, by Theorem \ref{realeigen}, $u$ has infinitely many zeros on the positive imaginary axis. 
  
The next theorem gives some information on how zeros of $u$ and $u^\d$ 
in $A_1$ should be arranged, when $\beta >0$. Note that all the zeros of $u$ and $u^\d$ in $\Im z \geq 0$ lie in $A_1$ by Theorem \ref{im-part}.
\begin{theorem} \label{locating_zeros_theorem}
Suppose $u(z)$ is an eigenfunction of (\ref{eigen2}) with eigenvalue
$\lambda \in \C$ with  $\Im \lambda=\beta > 0$. Then 
\item{(i)} $\Re (u^\d \bar u) \geq 0$ for some point on the imaginary axis if 
and only if $uu^\d$ has infinitely many zeros in $A_1 \cap B_2$ and
at most finitely many zeros in $A_1 \cap B_2^c$; and 
\item{(ii)} $\Re (u^\d \bar u) < 0$ for every point on the imaginary axis if  
and only if $uu^\d$ has no zeros in $ \{z \in A_1: \Re z \leq 0\}$ and 
infinitely many  in  $ \{z \in A_1: \Re z> 0\}$. 
\end{theorem}
We will use the following lemma along with Lemma \ref{mono_sign2}. 
\begin{lemma}\label{mono_sign1}
Assume $\Im \lambda=\beta > 0$. Suppose $\Re \zeta_1 \leq \Re \zeta_2$ and  
 $\Re (u^\d \bar u)=0$ at $\zeta_1$, $\zeta_2$ $(\text{where } \zeta_1 \not =\zeta_2)$. 
Then
\item{(i)} $\zeta_1, \zeta_2 \in cl(A_1 \cap B_2) \Longrightarrow$
$\Re \zeta_1 < \Re \zeta_2$ and $\Im \zeta_1 < \Im \zeta_2$, and 
\item{(ii)} $\zeta_1, \zeta_2 \in cl(A_1 \cap B_2^c) \Longrightarrow$
$\Re \zeta_1 < \Re \zeta_2$ and $\Im \zeta_1 > \Im \zeta_2$. 
\end{lemma}
\begin{proof}[Proof of part (i)] 
We will first prove this for  $\zeta_1, \zeta_2 \in A_1 \cap B_2$.
Suppose that $\Re \zeta_1 = \Re \zeta_2$. Then we could find a
vertical line segment $z(t)$ in $A_1 \cap B_2$ 
whose end points are $\zeta_1$ and $\zeta_2$. 
We apply (\ref{real-curve})  to this line segment to get
$$
0= -\int_c^d \Im (iz^3(t)-\lambda)|u(z(t))|^2 dt.
$$
This would imply $u \equiv 0$ on the curve $z(t)$ since $ \Im
(iz^3(t)-\lambda) >0$ in  $B_2$. So then since $u$ 
is analytic, $u \equiv 0$ in $\C$. This is a contradiction.
Hence $\Re \zeta_1 < \Re \zeta_2$.
 
Similarly, suppose that $\Im \zeta_1 \geq \Im \zeta_2$.
Then we could find a smooth curve $z(t)=x(t)+iy(t)$ in $A_1 \cap B_2$ 
such that $z(c)=\zeta_1$,  $z(d)=\zeta_2$, $x^\d(t)> 0$ and $y^\d(t)\leq 0$.
Note that $\Im (iz^3(t)-\lambda)>0$ and $\Re (iz^3(t)-\lambda)>0$ in
$A_1 \cap B_2$. This contradicts (\ref{real-curve}) like for the case
of $\Re \zeta_1 = \Re \zeta_2$.
We now see that the above argument still holds for 
$\zeta_1, \zeta_2 \in cl(A_1 \cap B_2)$.

\noindent {\it Proof of part (ii).} We use (\ref{real-curve}) again and a similar argument like in the proof of part $(i)$.
\end{proof}
\begin{figure}[t]
    \begin{center}
    \includegraphics[width=.4\textwidth]{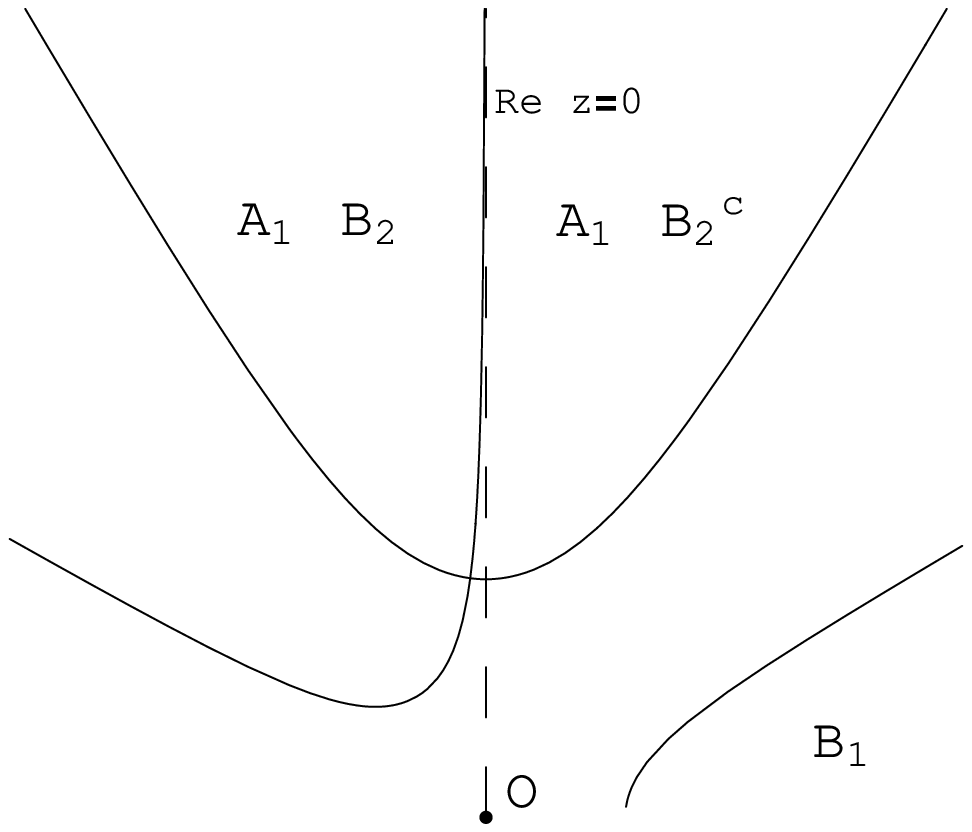}
    \end{center}
 \vspace{-.5cm}
\caption{  }\label{f:shin8}
\end{figure}
\begin{proof}[Proof of Theorem ~\ref{locating_zeros_theorem}]
 Suppose $u(z)$ is an eigenfunction of (\ref{eigen2}) with  eigenvalue
$\lambda \in \C$ with  $\Im \lambda=\beta > 0$. Since $u$ has infinitely many zeros in $A_1$ (by the paragraph shortly before Theorem \ref{locating_zeros_theorem}), certainly $uu^\d$ also has infinitely many zeros in $A_1$.

\noindent {\it Proof of part (i).} 
Suppose that  $\Re (u^\d \bar u) \geq 0$ for some point $iy$ on the
imaginary axis. By (\ref{real-curve}) with $z(t) = it$, we have that 
\begin{equation}\label{imaginary_real}
\left. \Re (u^\d \bar u)\right|_{ic}^{id}=\beta \int_c^d |u(iy)|^2 dy.
\end{equation}
So then since $\beta >0$, $\Re (u^\d \bar u)
> 0$ at every point $it$ for $t>y$. Now  by Lemma \ref{mono_sign2}, 
we have
that $\Re (u^\d \bar u) > 0$ at every $x+it \in A_1$ for $t>y$ and $x \geq 0$.
Thus $uu^\d$ does not have any zeros in $\{z \in A_1 \cap B_2^c :\Re z \geq 0, \Im z > y \}$.

The entire function $uu^\d$ does not have infinitely many zeros in any bounded region. So if 
$uu^\d$ had infinitely many zeros in $A_1 \cap B_2^c$, then $uu^\d$ would have
infinitely many zeros in $ \{z \in A_1 \cap B_2^c : \Re z<0\}$. But if 
$uu^\d$ has a zero $z_1$ in $ \{z \in A_1 \cap B_2^c:\Re z<0\}$,  then by
Lemma \ref{mono_sign1} $(ii)$, $uu^\d$ has no zeros in $\{z \in A_1 \cap B_2^c :  
\Re z_1 \leq \Re z<0\}$. So then $uu^\d$ would have infinitely many 
zeros in a bounded region. This is a  contradiction. 
Thus $uu^\d$ has infinitely many zeros in $A_1 \cap B_2$ and at most
finitely many zeros in $A_1 \cap B_2^c$. 

Conversely, suppose that  $uu^\d$ has infinitely many zeros in $A_1 \cap B_2$ 
and at most finitely many zeros in $A_1 \cap B_2^c$.
Choose a zero $z_0$ in $A_1 \cap B_2$. Then by Lemma \ref{mono_sign2},
 we see that $\Re (u^\d \bar u)> 0$ at $i\Im z_0$ since $\Re z_0 < 0$. 

\noindent {\it Proof of part (ii).} 
Suppose that $\Re (u^\d \bar u)< 0$ for every point on the imaginary
axis. Then by Lemma \ref{mono_sign2}, $\Re (u^\d \bar u)< 0$ for
every point in $\{z \in A_1: \Re z \leq 0\}$. So then $uu^\d$ has no zeros in
$\{z \in A_1: \Re z \leq 0\}$.   
Now since we know that $uu^\d$ has infinitely many zeros in $A_1$, $uu^\d$
must have infinitely many zeros in  $ \{z \in A_1: \Re z > 0\}$.

Conversely, suppose $\Re (u^\d \bar u) \geq 0$ 
for some point on the imaginary axis. Then $uu^\d$ would have
at most finitely many zeros in   $ \{z \in A_1: \Re z> 0\}$ by the
argument as in the proof of part $(i)$. This completes the proof.
\end{proof}
\begin{remark}
{\rm
 Since the negative imaginary axis is in the middle of  a blowing-up Stokes 
region (see Figure \ref{f:shin2}), $u(iy)$ blows up as $y$ tends to $-\infty$.
 On the other hand, the positive imaginary axis is a critical ray. We can show
 that 
$|u(iy)|^2 \leq (const.) y^{-{\frac{3}{2}}}$ for all $y$ near positive 
infinity, by  Theorem 7.4.4 in \cite{H}.

So the right-hand side of (\ref{imaginary_real}) approaches $+\infty$ as 
$c$ tends to $-\infty$ (while $d$ is fixed). Thus we 
see that  
$\Re [u^\d(ic) \bar u(ic)]< 0$  for all $c$ near negative infinity. 
However, the right-hand side of (\ref{imaginary_real}) is convergent as 
$d$ tends to $+\infty$ (while $c$ is fixed). So $\Re(u^\d \bar u)$ may or may
not become positive near infinity along the positive imaginary axis.} 
\end{remark}

The next lemma gives some information on  zeros of $u$ and $u^\d$ in 
$\Im z <0$, if any exist. There can only be finitely many such zeros, by the paragraph shortly before Theorem \ref{locating_zeros_theorem}.
\begin{lemma}\label{below=0}
Assume $\Im \lambda=\beta \geq 0$. Suppose $\Im \zeta_1 \leq \Im \zeta_2$ and  
 $\Im (u^\d \bar u)=0$ at $\zeta_1$, $\zeta_2$ $(\text{where } \zeta_1 \not 
=\zeta_2)$.  Then
\item{(i)} $\zeta_1, \zeta_2 \in cl(A_4 \cap B_3) \Longrightarrow$
$\Im \zeta_1 < \Im \zeta_2$ and $\Re \zeta_1 < \Re \zeta_2$, and 
\item{(ii)} $\zeta_1, \zeta_2 \in cl(A_4 \cap B_4) \Longrightarrow$
$\Im \zeta_1 < \Im \zeta_2$ and $\Re \zeta_1 > \Re \zeta_2$. 
\end{lemma}
\begin{proof}
We omit the proof because it is very similar to the proof of Lemma \ref{mono_sign1}.
We use (\ref{im-curve}) instead of (\ref{real-curve}), and also make use of Figures \ref{f:shin6} and \ref{f:shin7}.
\end{proof}

Roughly speaking, then, the zeros move up and to the right in the third 
quadrant, and down and to the right in the fourth quadrant.
This observation  supports that when $\lambda$ is real, zeros of $u$ 
in $\Im z <0$ lie on an arch-shaped curve as in Figures 5 and 6 in \cite{BBS}. 
\section{Other properties of eigenfunctions}\label{other}
Here we present a possible way of proving the conjecture that the
eigenvalues $\lambda$ of $H=-\frac{d^2}{dz^2}-(iz)^3$ are positive
real. Given an eigenfunction $u$ with eigenvalue $\lambda$,  
Theorem \ref{functions} below gives a class $\mathcal O$ of polynomials $p(x,y)$ 
which are orthogonal to $|u|^2$ in the sense that $\int_{-\infty}^{\infty}
p(x,y)|u(x+iy)|^2 dx=0$  for all $y$.
One can perhaps prove the conjecture as follows.
Suppose $\Im \lambda \not =0$; if $\mathcal O$ is large enough 
then $|u|^2 \equiv 0$, giving a contradiction.  

Let $u$ be an eigenfunction of  $H=-\frac{d^2}{dz^2}-(iz)^3$ with eigenvalue 
$\lambda=\alpha+i\beta$.
\begin{theorem} \label{functions}
Let $\mathcal{O}=\{\text{polynomials } p(\cdot, \cdot): \int_{-\infty}^{\infty}
p(x,y)|u(x+iy)|^2 dx=0 \text{ for 
all $y$}  \}$. Then:

\item{(i)} $x^3-3xy^2-\beta \in \mathcal O$,
\item{(ii)} for all $m \geq 0$,
\begin{eqnarray}
\frac{4}{m + 1}(\frac{ x^{m + 5} }{m + 5} -3y^2 
        \frac{ x^{m + 3}}{m + 3} -\beta \frac{x^{m + 2}}{m + 2})(x^3 - 
        3y^2 x - \beta)& &\nonumber\\
 - m(m -1)x^{m-2} -  4x^m(3x^2 y - y^3 + \alpha) - \frac{12}{m + 1}y
        x^{m + 2} &\in& \mathcal O,\nonumber 
\end{eqnarray}
\item{(iii)} if $ p \in \mathcal O$ then
$p_y+2(x^3-3xy^2-\beta)\int_0^xp(t,y)  
dt \in \mathcal O$, and
\item{(iv)} if $p \in \mathcal O$ then
$p_{xx}+p_{yy}+12x^2yp+4(x^3-3xy^2-\beta)\int_0^x p_y(t,y) dt
\in \mathcal O$.
\end{theorem}
For example the following polynomials are in $\mathcal O$:
\begin{eqnarray}
p_3(x,y) &=& x^3-3xy^2-\beta, \text{  by $(i)$},\nonumber\\
p_7(x,y) &=& x^7-9x^5y^2-5x^4\beta+18x^3y^4+18x^2y^2\beta
+4x(\beta^2-3y),\nonumber\\
& &\text{by applying  $(iii)$ to $p_3$ and multiplying by 2},\nonumber\\
p_8(x,y) &=& 2x^8-16x^6y^2-7x^5\beta+30x^4y^4+25x^3y^2\beta
+5x^2(\beta^2-12y)+10y^3-10\alpha,\nonumber\\
& & \text{by  applying $(ii)$ with $m=0$ and multiplying by $\frac{5}{2}$,}\nonumber\\
p_9(x,y) &=& 2x^9-15x^7y^2-6x^6\beta+27x^5y^4+
4x^3(\beta^2-27y)+24xy^3+21x^4y^2\beta-24x\alpha, \nonumber\\
& &\text{by  applying $(ii)$ with
$m=1$ and multiplying by $6$, and}\nonumber\\
p_{10}(x,y) &=&
20x^{10}-144x^8y^2-55x^7\beta+252x^6y^4+189x^5y^2\beta-35x^4(48y-\beta^2)
\nonumber\\  
& &+420x^2(y^3-\alpha)-210, \text{ by applying $(ii)$ with $m=2$ and multiplying  by $105$.}\nonumber 
\end{eqnarray}
We do not know whether Theorem \ref{functions} generates all the polynomials in $\mathcal O$.
\begin{proof}[Proof of Theorem ~\ref{functions}]
It is useful to have the following two formulas, which follow from 
multiplying (\ref {eigen2}) by $\bar u$ and
separating real and imaginary parts: 
\begin{equation} \label{im-diff}
 \Im [u_x(x+iy)\bar u(x+iy)]_x=(x^3-3xy^2-\beta)|u(x+iy)|^2 
\end{equation}
and
\begin{equation} \label{real-diff}
\Re [u_x(x+iy)\bar u(x+iy)]_x=|u_x|^2+(-3x^2y+y^3-\alpha)|u|^2.
\end{equation}
At the end of the proof we will justify the fact that we can
differentiate through the integrals that follow. 

$(i)$ This is clear by integrating (\ref{im-diff}), using the zero boundary conditions in the left- and right-hand Stokes regions.

$(ii)$ Suppose $m$ is a nonnegative integer. Then

\begin{eqnarray}
\frac{d}{dy}\int_{-\infty}^{\infty} x^m|u|^2
dx=\int_{-\infty}^{\infty} x^m \frac{\partial}{\partial y}|u|^2
dx&=&-2\int_{-\infty}^{\infty} x^m \Im (u_x \bar u) dx \label{eq_1} \\
&=&\frac{2}{m+1}\int_{-\infty}^{\infty} x^{m+1}\Im (u_x \bar u)_x dx,
\nonumber
\end{eqnarray}  
where the last step is by integration by parts.

 So using (\ref{im-diff}), we have that
\begin{equation} \nonumber
\frac{d}{dy}\int_{-\infty}^{\infty}
x^m|u|^2dx=\frac{2}{m+1}\int_{-\infty}^{\infty}
x^{m+1}(x^3-3xy^2-\beta)|u|^2 dx. 
\end{equation}

Hence,
\begin{eqnarray}
 & &\frac{d^2}{dy^2}\int_{-\infty}^{\infty} x^m|u(x+iy)|^2 dx \nonumber\\ 
&=&\frac{2}{m+1}\int_{-\infty}^{\infty}
[-6x^{m+2}y|u|^2-2(x^{m+4}-3x^{m+2}y^2-\beta x^{m+1})\Im (u_x \bar u)]
dx.\nonumber
\end{eqnarray}

Then again using  the integration by parts and (\ref{im-diff}), we
have that this equals 
\begin{equation}\label{eq_4}
\frac{2}{m+1}\int_{-\infty}^{\infty} [-6x^{m+2}y+2(\frac{x^{m+5}}
{m+5}-3y^2\frac{x^{m+3}}{m+3}-\beta \frac{ x^{m+2}}{m+2}) 
(x^3-3xy^2-\beta)]|u|^2 dx.
\end{equation}

Also, we differentiate (\ref{eq_1}) without applying integration by
parts:
\begin{eqnarray}
\frac{d^2}{dy^2}\int_{-\infty}^{\infty} x^m|u|^2
dx&=&-2\int_{-\infty}^{\infty} x^m [\Re (u_x \bar u)_x-2|u_x|^2] dx 
\label{eq_5}\\
&=&-2\int_{-\infty}^{\infty} x^m[(-3x^2y+y^3-\alpha)|u|^2-|u_x|^2] dx
\quad \text{by (\ref{real-diff})}.\label{eq_6}
\end{eqnarray}

Also, applying integration by parts twice to the right-hand side of
(\ref{eq_5}),  we have that (\ref{eq_5}) equals 
\begin{equation} \label{eq_7}
-m(m-1)\int_{-\infty}^{\infty} x^{m-2}|u|^2
 dx+4\int_{-\infty}^{\infty} x^m|u_x|^2 dx. 
\end{equation}
By equating (\ref{eq_6}) and  (\ref{eq_7}), we get 
\begin{equation} \label{eq_8}
\int_{-\infty}^{\infty} x^m|u_x|^2
dx=\frac{m(m-1)}{2}\int_{-\infty}^{\infty} x^{m-2}|u|^2
dx-\int_{-\infty}^{\infty} 
x^m(-3x^2y+y^3-\alpha)|u|^2 dx.
\end{equation}
Hence equating (\ref{eq_4}) and (\ref{eq_6}) and substituting (\ref{eq_8})
 give $(ii)$.

$(iii)$ Suppose that $\int_{-\infty}^{\infty} p(x,y)|u|^2 dx =0 \text{
for all $y$}$. Then
\begin{eqnarray} 
0&=&
\frac{d}{dy}\int_{-\infty}^{\infty} p(x,y)|u|^2 dx \nonumber\\
&=&
\int_{-\infty}^{\infty} [p_y(x,y)|u|^2 -2 p(x,y) \Im (u_x \bar u)] dx
\label{eq_100}\\
&=&
\int_{-\infty}^{\infty} \left[p_y(x,y)|u|^2+2\biggl\{ 
\int_0^x p(t,y)dt\biggr\} \Im (u_x \bar u)_x \right] dx,\nonumber
\end{eqnarray}
 by integration by parts. This with (\ref {im-diff}) gives $(iii)$.

$(iv)$ Suppose $\int_{-\infty}^{\infty} p(x,y)|u|^2 dx =0 \text{ for all }
y$.

Then we differentiate through (\ref{eq_100}) with respect to $y$ again
to get
\begin{eqnarray} 
0&=&\int_{-\infty}^{\infty} \left[p_{yy}(x,y)|u|^2-4p_y(x,y)\Im (u_x \bar
u)-2 p(x,y)\Im (iu_{xx} \bar u-i|u_x|^2)\right] dx \nonumber\\
&=&\int_{-\infty}^{\infty}
\left[p_{yy}(x,y)|u|^2+4\biggl\{\int_0^xp_y(t,y)dt\biggr\}\Im (u_x \bar 
u)_x-2 p(x,y)(\Re (u_x \bar u)_x-2|u_x|^2)\right] dx\nonumber\\
&=&\int_{-\infty}^{\infty}
\left[p_{yy}(x,y)+4\biggl\{\int_0^xp_y(t,y)dt\biggr\}(x^3-3xy^2-\beta)\right]
|u|^2 dx\nonumber\\
& &-2
\int_{-\infty}^{\infty}p(x,y)\left[-\Re (u_x \bar
u)_x+2(-3x^2y+y^3-\alpha) |u|^2\right] dx, \qquad \text{by (\ref{im-diff}) and (\ref{real-diff})}\nonumber\\
&=&\int_{-\infty}^{\infty}
\left[p_{yy}(x,y)+4\biggl\{\int_0^xp_y(t,y)dt\biggr\}(x^3-3xy^2-\beta)-4p(x,y)
(-3x^2y+y^3-\alpha)\right]|u|^2
dx\nonumber\\
& &-\int_{-\infty}^{\infty} p_x(x,y) 2 \Re (u_x \bar u) dx.\nonumber
\end{eqnarray}
But $(y^3-\alpha)\int_{-\infty}^{\infty}p|u|^2 dx=0$, 
and so applying integration by parts again to the last term gives $(iv)$.

Now to complete the proof we need to show that we can differentiate
through the above integrals, which reduces to showing that  
$$
\frac{d}{dy}\int_{-\infty}^{\infty} x^m|u|^2 dx=\int_{-\infty}^{\infty} x^m
\frac{\partial}{\partial y}|u|^2 dx, \qquad \text{for each $m\geq 0$}.
$$
 So we estimate the following:
\begin{eqnarray}
& &\biggl| \int_{-\infty}^{\infty} x^m[ \frac{1}{h}\{|u(x+i(y+h))|^2-
|u(x+iy)|^2\}-\frac{\partial}{\partial y}|u(x+iy)|^2] dx
\biggr|\nonumber\\ 
& \leq & \int_{-\infty}^{\infty}|x|^m\bigl|
\frac{1}{h}\{|u(x+i(y+h))|^2-|u(x+iy)|^2\}-\frac{\partial}{\partial
y}|u(x+iy)|^2\bigr| dx \nonumber\\ 
& \leq &|h|
\int_{-\infty}^{\infty}|x|^m\bigl|\frac{\partial^2(|u|^2)}{\partial
y^2}(x+i\xi(x)) \bigr| dx, \quad \text{by the Mean Value Theorem,}
\nonumber 
\end{eqnarray}
where $|\xi(x)-y|\leq |h|.$

So then it is enough to show that for each $M>0$, there exist $C_1>0$
and $C_2>0$ such that  
\begin{equation}\label{ability}
|u(x+iy)|+|u^\d(x+iy)|+|u^\dd(x+iy)| \leq \frac{C_1}{\exp[|x|^{C_2}]}
 \qquad \text{for all} \quad |y|\leq M. 
\end{equation}
Proof of (\ref{ability}) is as follows:

Suppose that $z=r e^{i\theta}$ with $|\theta| \leq\frac{\pi}{20}$. Then $z$ 
is in the decaying Stokes regions (see Figure \ref{f:shin2}). 
By the asymptotic expression (\ref{asymp-formula}) we get that for
some $C >0$, $|u(re^{i\theta})| \leq \exp[-|r|^{C}]$ for all
$|\theta|\leq  \frac{\pi}{20}$ and large $|r|$, say $|r| \geq R$.  Also
choose $R \geq M+1$.  

Now the Cauchy integral formula  says that
$$
u^{(k)}(z)= \frac{k!}{2\pi
i}\int_{|\zeta-z|=1}\frac{u(\zeta)}{(\zeta-z)^{k+1}} d\zeta.$$ 
So then
$$|u^{(k)}(z)| \leq \frac{k!}{2\pi}\int_{|\zeta-z|=1} |u(\zeta)| |d\zeta|
\leq k! \max\{|u(\zeta)| : |\zeta - z|=1 \} \leq k! \exp[-(|z|-1)^{C}]$$
 where the last inequality holds if $\{\zeta \in \C: |\zeta-z|=1\} 
\subset \{\zeta \in \C: \zeta= \rho e^{i\phi}, |\rho| \geq R, |\phi|\leq \frac{\pi}{20}\}$.  

Choose $0 < C_2 < C$. Then $\exp[-(|z|-1)^{C}] \leq \exp[-|z|^{C_2}]$ if 
$|z|\geq R_2$ for large $R_2\geq R$. 

The region where $|z|\geq R_2$ and $\{\zeta \in \C : |\zeta-z|=1\} 
\subset \{\zeta \in \C: \zeta= \rho e^{i\phi}, |\rho| \geq R, |\phi|\leq \frac{\pi}{20}\}$ 
covers  all of $|y|\leq M$ but a bounded region.
  
Since the minimum of  $\exp[-|z|^{C_2}]$ in this bounded region
is strictly positive, we can find a large $C_1>0$ so that the left-hand 
side of  
(\ref{ability}) is bounded by $C_1 \exp[-|z|^{C_2}]$. 
Thus (\ref{ability})  holds since $|x| \leq |z|$.
And this completes the  proof. 
\end{proof}
\begin{corollary} \label{convex}
 Let $u(z)$ be an eigenfunction of (\ref {eigen2}). Then
$\int_{-\infty}^{\infty} |u(x+iy)|^2 dx$ is a convex function.
\end{corollary}
\begin{proof}
This is a consequence of (\ref{eq_7}) with $m=0$, or it can be proved
using the subharmonicity of $|u|^2.$ 
\end{proof}
\section{Conclusions}
\label{conclusions}
Using simple path integrations, we were able to prove that eigenvalues
of (\ref{eigen}) lie in the sector  $|\arg \lambda| \leq
\frac{\pi}{2n+3}$ and we extended the result for some more general
Hamiltonians.    Also we provide  zero-free regions of eigenfunctions and
their first derivatives, for the potential $-(ix)^3$. Then finally we
have the  set $\mathcal O$ of polynomials $p(x,y)$ which are
orthogonal to $|u|^2$ in the sense that $\int_{-\infty}^{\infty}
p(x,y)|u|^2 dx=0$ for  all $y$. 

In a recent communication with Mezincescu, he pointed out that for the 
potential $-(ix)^3$ if $\Im \lambda=\beta \not =0$, combining $\frac{|\beta|}{\alpha} \leq \tan \frac{\pi}{5}$ with the equation ($23$) in \cite{M} gives $|\lambda|>(\frac{2}{5} \cos\frac{\pi}{5}) 10^5\approx 3 \times  10^4.$ So if any non-real eigenvalues exist, they are very large.
  
In this paper we consider only polynomial potentials with odd degrees. However, a number of other 
authors have worked on even degree potentials, particularly quartic \cite{CD, MO} and sextic \cite{B, BCQ} polynomial 
potentials. Our techniques in proving Theorems \ref{sector-theorem} and \ref{extended_1} can be 
used to get information on eigenvalues for even degree potentials if both ends of a line passing through the origin stay in decaying Stokes regions. 
 
Obvious open problems are to narrow the eigenvalue sectors closer to the positive
real axis, and finally to prove that the eigenvalues are real. Since some
$\mathcal  {PT}$-symmetric non-Hermitian Hamiltonians do not have all
real eigenvalues, one might further want to classify $\mathcal
{PT}$-symmetric  non-Hermitian Hamiltonians which do have positive real
eigenvalues. 
\subsection*{{\bf Acknowledgments}}

The author was partially supported by NSF grant number
DMS-9970228. The author appreciates Gary Gundersen's help at the
initial stage and Carl Bender's comments later on, and thanks Richard
S. Laugesen for encouragement, invaluable suggestions and discussions
throughout the work.      
\vskip12pt

{\sc email contact:}  kcshin@math.uiuc.edu
\end{document}